\documentclass[journal]{IEEEtran}

\usepackage{graphicx}
\usepackage{amsmath}
\usepackage{cite}
\usepackage{multirow}
\usepackage{amsfonts,amssymb}
\usepackage{flushend, cuted}
\usepackage{lipsum}
\usepackage{stfloats}
\usepackage{cases}
\usepackage{multirow}
\usepackage{booktabs}
\usepackage{balance}
\usepackage{enumitem}
\usepackage{inconsolata}
\usepackage{bbm}
\usepackage{color}
\usepackage{makecell}
\usepackage{url}
\usepackage[ruled,linesnumbered]{algorithm2e}
\ifCLASSINFOpdf

\else

\fi
\begin{document}
\title{\LARGE Electrical Behavior Association Mining for Household Short-Term Energy Consumption Forecasting}
\author{Heyang~Yu,~\IEEEmembership{Graduate~Student~Member,~IEEE},
	           Yuxi~Sun,~Yintao~Liu,\\
			   Guangchao~Geng,~\IEEEmembership{Senior~Member,~IEEE},
       		   and~Quanyuan~Jiang,~\IEEEmembership{Senior~Member,~IEEE}
\thanks{H. Yu, Y. Sun, G. Geng and Q. Jiang are with the College of Electrical Engineering, Zhejiang University, Hangzhou 310027, China. (e-mail: $\{$yuheyang, sunyuxi, ggc, jqy$\}$@zju.edu.cn). Y. Liu is with the Polytechnic Institute, Zhejiang University, Hangzhou 310015, China. (e-mail: 22160186@zju.edu.cn). \emph{Corresponding author: Guangchao Geng}.}
	%Y. Sun are with the Institute of Economics and Technology State Grid Zhejiang Electric Power Co. Ltd., Jinan, 250 021, China (e-mail: sunyuxi@zju.edu.cn). 
	}
% The paper headers
\markboth{Submitted to Journal for possible publication}%
{Shell \MakeLowercase{\textit{et al.}}: Bare Demo of IEEEtran.cls for IEEE Journals}
\maketitle
\begin{abstract}
Accurate household short-term energy consumption forecasting (STECF) is crucial for home energy management, but it is technically challenging, due to highly random behaviors of individual residential users. To improve the accuracy of STECF on a day-ahead scale, this paper proposes an novel STECF methodology that leverages association mining in electrical behaviors. First, a probabilistic association quantifying and discovering method is proposed to model the pairwise behaviors association and generate associated clusters. Then, a convolutional neural network-gated recurrent unit (CNN-GRU) based forecasting is provided to explore the temporal correlation and enhance accuracy. The testing results demonstrate that this methodology yields a significant enhancement in the STECF.
\end{abstract}

\begin{IEEEkeywords}
Association mining, electrical behavior, short-term energy consumption forecasting, spectral clustering.
\end{IEEEkeywords}
\IEEEpeerreviewmaketitle
%\section*{Nomenclature} %术语表，打星号是因为不用参与正文标题的标号排序
%\subsection*{Indices} % 小标题，可以分为集合，下标等等
%\begin{IEEEdescription}[\IEEEusemathlabelsep\IEEEsetlabelwidth{$i, j, x, y$}] % 设置符号的最大宽度
%	\item[$i, j, x, y$] Indices for nodes.
%	\item[$h$] Index for the load levels used to represent the loading condition.
%\end{IEEEdescription}
\vspace{-1.0em}
\section{Introduction}
\IEEEPARstart{A}{ccurate} household short-term energy consumption forecasting (STECF) plays a pivotal role in household energy management system (HEMS), managing the efficiency of power generation components, and energy storage systems \cite{8039509}, which also contributes to the attainment of energy-saving and carbon emission reduction. However, the residential load is of small scale, meanwhile with the characteristics of diverse types and quantities of appliances, random user's behavior, and being greatly affected by weather, special activities, etc., which imposes challenges to individual residential STECF.

The widespread implementation of smart meter infrastructure has opened up opportunities for STECF to tackle the aforementioned challenges. Some researches focus on exploring data-driven methods to improve the accuracy of household STECF. Long and short-term memory network (LSTM) \cite{8039509, WANG201910, 9917171} and gated recurrent unit (GRU) \cite{WANG2021107233} approach have been proposed to enhance the accruacy of household STECF. Nevertheless, fine-grained energy consumption data can be used for this work. Ref. \cite{7894203} has revealed energy consumption patterns through frequent pattern mining using time series data from various electrical appliances. Ref. \cite{8933148} proposes a conditional hidden semi-Markov model to describe the probabilistic nature of residential appliance demand and related load forecasting algorithms. However, they both neglect the relationship between different electrical appliances. In Ref. \cite{8606124, 9102266}, researchers have proposed a non-intrusive load monitoring (NILM) and graph spectral clustering-based method that utilize the correlations of appliances behaviors in terms of their state (ON/OFF) durations. Ref. \cite{chen2023graph} proposes graph-based method to represent the behavior in load identification and energy management. Nevertheless, the aforementioned researches primarily focus on intra-day and ultra-short-term load prediction. These methodologies are not applicable for individual residential forecasting on the day-ahead scale. %As a result, a day-head short-term energy consumption forecasting method that comprehensively considers the different types of electrical behavior associations to address the challenges faced by household STECF. 

In light of the above discussion, to tackle multiple challenges faced by household STECF, this paper proposes an day-head STECF method ensembling association in pairwise appliances and their temporal correlation. First, a probabilistic association quantifying and appliance aggregation method is proposed to model the pairwise appliances consumption association and generate associated clusters. Then, a convolutional neural network (CNN)-GRU based forecasting method is provided to explore the temporal correlation and enhance accuracy. The results show that this method achieves an improvement in the performance on the day-head scale.  Contributions of this work include:

(i) A probabilisitc quantification method using improved graph-based representation is proposed, which fully utilizes the association between pairwise appliances and improves day-ahead predictability. 

(ii) Based on the appliance behavior related clusters, the proposed forecasting model can be lightweight  simultaneously than overall forecasting without significantly losing prediction accuracy.
\vspace{-1.0em}
\section{Behavior Association Mining Assisted Forecasting}\label{2}
%In this section, a probabilistic quantification and aggregation method for appliances is used to describe and quantify the association between pairwise electrical behaviors (such as consumption time) mined from fine-grained historical electrical behavior data, and to aggregate electrical appliances with strong association to a cluster.
\subsection{Probabilistic Quantification of Electrical Behavior Association}\label{2.1}
In Ref. \cite{chen2023graph}, the behavior association can be expressed by the probability that an appliance is used after another appliance. In this work, we extend this idea that behavior association is characterized as the probability of an appliance being used before or after another appliance (both appliances are working) for a duration of $T_e$ which is the target time for the association. The behavior association matrix $\boldsymbol{Q}$ can be used to describe the probability of pairwise behavior association among a group of appliances, which can be expressed as  (\ref{eq2023121801}):
\begin{equation}
	\setlength{\abovedisplayskip}{3pt} %%% 3pt 个人觉得稍妥，可自行设置
	\setlength{\belowdisplayskip}{3pt}
	\left\{
	\begin{aligned}
		&\boldsymbol{Q} : \{Q_{i, j}: 1 \leq i ,j \leq N^{\text{a}}\}\\
		&Q_{i, j} = {Q(a_i \cdot a_j | T_e) \over Q(a_i \cdot a_j | T_s)} \cdot {N^{\text{d}}_{i,j}\over N^{\text{d}}}= {N^{\text{e}}_{i,j}\over N^{\text{s}}_{i,j}} \cdot {N^{\text{d}}_{i,j}\over N^{\text{d}}}
	\end{aligned}
	\right.
	\label{eq2023121801}
\end{equation}
where $Q_{i, j}$ is the behavior association probability that appliance $a_j$ and appliance $a_i$ are on for a duration of $T_e$. $Q(a_i \cdot a_j | T_e)$ and $Q(a_i \cdot a_j | T_s)$ are respectively the probability of both $a_i$ and $a_j$ being turned on during $T_e$ and $T_s$. $T_s$ is the time that can be used as a valid candidate. $N^{\text{a}}$ represents the total number of appliances. $N^{\text{e}}_{i, j}$ and $N^{\text{s}}_{i, j}$ are respectively the total number of both $a_i$ and $a_j$ being turned on during $T_e$ and $T_s$ in the time series data. $N^{\text{d}}_{i,j}$ and $N^{\text{d}}$ are number of days that both appliances have been used and total number of days.

It should be noted that appliance event and consumption data can be obtained via NILM modules in smart meters \cite{10147903}. The behavioral association is calculated using start-up time for its convenience where closing or consumption duration time can also be used. Additionally, the diagonal elements of $\boldsymbol{Q}$ are set to 0, eliminating interference during subsequent clustering.
\begin{figure}[t] %可在该句后⾯加上[h]，[t]，[b]等语句控制图⽚位置，其中[h]
	%表示原位置，[t]和[b]分别表示该⻚的顶端和底端
	\centering %居中
	\includegraphics[width=0.8\linewidth]{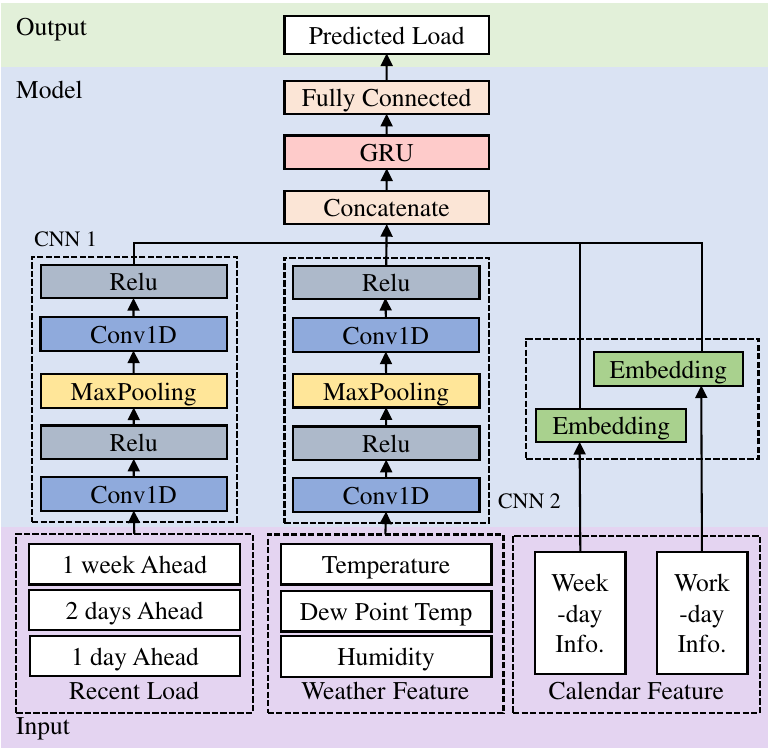} %定义图⽚宽度：0.25in
	\caption{Model structure of household STECF} %图⽚的标题内容
	\label{Network} %图⽚的标签，⽤于正⽂中引⽤该图⽚
	\vspace{-1.0em}
\end{figure}
\vspace{-2.0em}
\subsection{Spectral Clustering for Electrical Appliance Clustering}\label{2.2}
The objective of extracting behavioral association is to amalgamate appliances demonstrating strong associative behavior, thereby forming appliance clusters. The intention of STECF for these clusters on day-ahead scale is to jointly predict the appliances involved in a single electricity consumption event, thereby weaken the impact of the diversity and abundance of appliance types. Concurrently, the behavioral association probability matrix reveals its intrinsic nature as a fully connected graph, wherein appliances serve as nodes and the association between pairswise appliances act as edges. This matrix essentially functions as the adjacency matrix of the graph, with its values indicating the degree of pairwise association amongst all appliances. 

Therefore, in this work, spectral clustering, which evolved from graph theory, is introduced to cut the graph, and the cut subgraphs are referred to as appliance clusters. Because the above section has already obtained the behavioral association probability matrix representing the degree of pairwise association, this work directly uses it to achieve spectral clustering. Moreover, silhouette coefficient and elbow rule can both be used to determine the optimal number of appliance clusters. 
\begin{figure}[t] %可在该句后⾯加上[h]，[t]，[b]等语句控制图⽚位置，其中[h]
	%表示原位置，[t]和[b]分别表示该⻚的顶端和底端
	\centering %居中
	\includegraphics[width=1.0\linewidth]{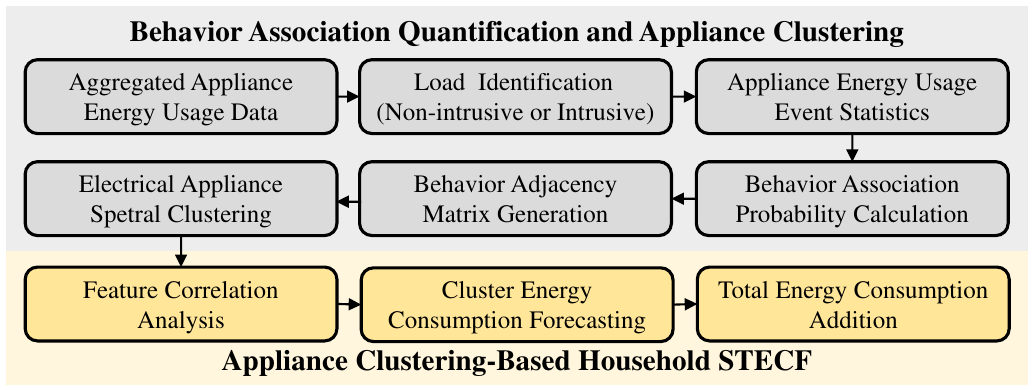} %定义图⽚宽度：0.25in
	\caption{The flowchart of the proposed method} %图⽚的标题内容
	\label{Flowchart} %图⽚的标签，⽤于正⽂中引⽤该图⽚
	\vspace{-1.0em}
\end{figure}
\begin{figure}[t] %可在该句后⾯加上[h]，[t]，[b]等语句控制图⽚位置，其中[h]
	%表示原位置，[t]和[b]分别表示该⻚的顶端和底端
	\centering %居中
	\includegraphics[width=0.5\linewidth]{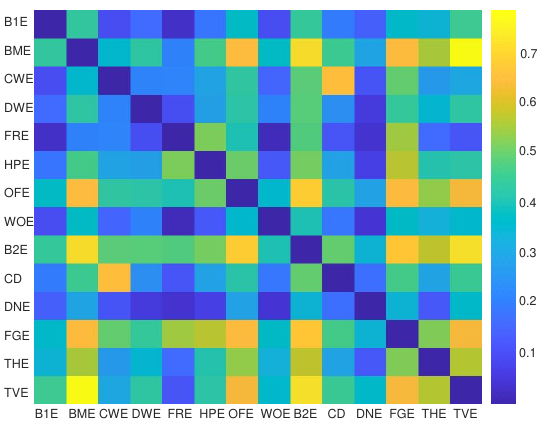} %定义图⽚宽度：0.25in
	\caption{Adjacency matrix (assocation probability matrix) of pairwise appliances} %图⽚的标题内容
	\label{Association} %图⽚的标签，⽤于正⽂中引⽤该图⽚
	\vspace{-1.0em}
\end{figure}
\vspace{-1.0em}
\subsection{Appliance Clustering-Based Household STECF}\label{2.3}
Due to the numerous features in household STECF and the strong nonlinearity of residential energy consumption, statistical learning is difficult to describe the actual input-output relationship. Therefore, after constructing the mutual dependence between pairwise appliances, deep learning is used to model the time-dependent use of electrical clusters.

Three types of input features are used: historical load, weather, and calendar features. Historical loads represent valuable past consumption patterns, weather represents the overall load consumption level of day, and calendar features represent the characteristics of human activities (work or vacation). Correlation analysis between features and load through distance correlation coefficient (DCC) \cite{WANG2021107233} is performed to achieve input feature selection. Consequently, this paper has designed a forecasting model, fundamentally built on CNN and GRU, as illustrated in Fig. \ref{Network}. The output is load of each time slot on day-ahead scale.

The function of convolutional layers is to deeply extract the characteristics of input variables, while GRU is to process long-term dependencies in sequence data. To handle heterogeneous input variables, calendar features  are encoded through an embedding layer before the convolutional layer. Meanwhile, to improve the effectiveness of prediction and prevent overfitting, the above features are normalized through min-max normalization. As a brief summary, the flowchart of the proposed method is provided as shown in Fig. \ref{Flowchart}.%\vspace{-1.0em}
\vspace{-1.0em}
\section{Case Study and Analysis}\label{4}
\begin{table}[t]
	\renewcommand{\arraystretch}{1.2}
	\setlength\tabcolsep{2pt}%调列距
	\caption{Parameter Setting of the Proposed Model}
	\vspace{-1.5em}
	\begin{center}
		\begin{tabular}{c c c c}
			\hline
			\hline
			Layer &Parameters &Layer &Parameters \\
			\hline
			\multirow{2}{*}{\makecell[c]{Conv1D\\ (CNN 1,2)}}
			&\multirow{2}{*}{\makecell[c]{Filters1=16\\  Filters2=64}}
			&\multirow{2}{*}{\makecell[c]{Conv1D \\ (CNN 1,2)}}
			&\multirow{2}{*}{\makecell[c]{Filters1=32\\  Filters2=128}}\\
			&&&\\
			GRU  &Units=32,  Layers=1 &GRU  &Units=64,  Layers=3\\
			\hline 
			\hline
		\end{tabular}
		\label{table3}
	\end{center}
	\vspace{-1.5em}
\end{table}
\begin{table}[t]
	\renewcommand{\arraystretch}{1.2}
	\setlength\tabcolsep{3pt}%调列距
	\caption{Appliance Clustering Results}
	\vspace{-1.5em}
	\begin{center}
		\begin{tabular}{c c c c}
			\hline
			\hline
			No.	&Electrical Appliances &No.	 & Electrical Appliances\\
			\hline
			1   &CD, CWE 		&3   &DNE\\
			\multirow{2}{*}{\makecell[c]{2}}
			&\multirow{2}{*}{\makecell[c]{B1E, BME, DWE, WOE\\OFE, B2E, THE, TVE, FGE}}
			&\multirow{2}{*}{\makecell[c]{4}}
			&\multirow{2}{*}{\makecell[c]{FRE, HPE}}\\
			&			&  &\\
			\hline
			\hline
		\end{tabular}
		\label{table1}
	\end{center}
	\vspace{-1.5em}
\end{table}
\subsection{Data Preparation and Experiment Setup}\label{4.1}
Three experiments are conducted utilizing the public dataset AMPds2 \cite{makonin2016ampds} to analyze the association between appliances, the correlation between loads and features, and to validate the effectiveness and advantages of the proposed STECF method. The dataset, spanning 730 days with 1-minute granularity main and sub-meter readings, and  1-hour granularity weather features, is ideal for algorithm testing due to its extensive timeframe and fine granularity. The training dataset comprises 23 months of data, with the remaining data used for testing the forecast model. Meanwhile, the forecasting target is day-ahead load (12 points, 2-hour time slot). As a regression problem, root mean square error (RMSE) and mean absolute error (MAE) are used to quantify the prediction error. The hyperparameters of the model are summarized in Table \ref{table3}. 
\vspace{-1.0em}
\subsection{Result of Association Mining Between Pairwise Appliances}\label{4.3}
In the dataset containing 18 appliances, 4 appliances with lower power ($<$50W) that contribute less to total energy consumption were excluded from the steps of association mining. Fig. \ref{Association} and Table \ref{table1} respectively show the probability matrix and aggregation results of pairwise appliances. It can be seen that there is a high correlation between appliances of the same category, while their correlation with other categories in electrical behavior is relatively low.
\vspace{-1.0em}
\subsection{Result of Feature Correlation Analysis}\label{4.4}
Table \ref{table2} reveals the DCC between four clusters and input features, demonstrating the effectiveness of association mining-based appliance aggregation. The DCC of clusters 2 and 4 (including most appliances) is comparable to or exceeds the DCC of total load, implying enhanced prediction outcomes. The low DCC of cluster 3, containing a single appliance, poses no significant impact. In cluster 1, appliances such as washing machines and dryers, which operate intermittently and for extended durations, contribute to a low DCC between current input features and load on the day-ahead scale. This indicates a fundamental limitation affecting the accuracy of small user load forecasts in the day-ahead timeframe.
\vspace{-1.0em}
\subsection{Effectiveness and Advantages of Proposed Method}\label{4.5}
The method proposed in this paper is compared with the overall forecasting method which directly predicts the total individual residential energy consumption where the input is recent load, weather feature and calendar feature as well as output is total load on the next day. The results in Table \ref{table4} displays the superior performance of our proposed method in day-ahead forecasting, outperforming overall load predictions by improving RMSE and MAE by 9.58\% and 14.44\% respectively. The number of parameters such as neurons and layers in the cluster prediction model should also be less than that of the overall prediction model. Here, 4 lower power appliances are incorporated into cluster 3. This enhancement is attributed to our method's clustering of appliances with similar behaviors, thereby improving load predictability. We also identified the inherent limitation in day-ahead energy consumption forecasting, stemming from the inability of current input features to predict the operation of certain appliances (such as appliances in cluster 1).
\begin{table}[t]
	\renewcommand{\arraystretch}{1.2}
	\setlength\tabcolsep{2pt}%调列距
	\caption{Distance Correlation Coefficient Between Load and Input Features}
	\vspace{-1.5em}
	\begin{center}
		\begin{tabular}{r c c c c c}
			\hline
			\hline
			&Total Load&Cluster 1&Cluster 2&Cluster 3 &Cluster 4\\
			\hline
			1 Week Ahead       &0.422 &0.280 &0.605 &0.173 &0.354\\
			2 Days Ahead        &0.435 &0.251 &0.622 &0.105 &0.414\\
			1 Days Ahead  		&0.462 &0.238 &0.615 &0.381 &0.455\\
			Temperature			&0.360 &0.287 &0.363 &0.262 &0.312\\
			Humidity         		&0.353 &0.271 &0.358 &0.253 &0.318\\
			Dew Point Temp   &0.366 &0.294 &0.360 &0.211 &0.318\\
			\hline
			\hline
		\end{tabular}
		\label{table2}
	\end{center}
	\vspace{-1.5em}
\end{table}
\begin{table}[t]
	\renewcommand{\arraystretch}{1.2}
	\setlength\tabcolsep{4pt}%调列距
	\caption{Forecasting Performance Comparison of Different Methods}
	\vspace{-1.5em}
	\begin{center}
		\begin{tabular}{c c c c c}
			\hline
			\hline
			Item                      &RMSE &RMSE &MAE &\%MAE \\
			\hline
			Overall Forecasting   &462.39 &$\sim$ &349.90&$\sim$  \\
			\multirow{2}{*}{\makecell[c]{Appliance Cluster Forecasting\\(Proposed Method)  }}&\multirow{2}{*}{\makecell[c]{418.09}}&\multirow{2}{*}{\makecell[c]{-9.58\%}}&\multirow{2}{*}{\makecell[c]{299.37}}&\multirow{2}{*}{\makecell[c]{-14.44\%}}\\
			&&&&\\
			\hline
			\hline
		\end{tabular}
		\label{table4}
	\end{center}
	\vspace{-1.5em}
\end{table}%
\vspace{-1.0em}
\section{Conclusion}\label{5}
This paper proposes a day-head short-term energy consumption forecasting method ensembling association in pairwise appliances and the temporal correlation, which achieves highly accurate prediction performance. We discover the inherent limitation of low performance for small-scale users in day-ahead predictions, which may be mitigated on intraday scales. Future research could explore sustainable updates to the association probability matrix and corrective methods for intraday energy consumption forecasting.%
\bibliographystyle{IEEEtran}
\bibliography{IEEEabrv,IEEEexample}

% Generated by IEEEtran.bst, version: 1.12 (2007/01/11)
\begin{thebibliography}{10}
\providecommand{\url}[1]{#1}
\csname url@samestyle\endcsname
\providecommand{\newblock}{\relax}
\providecommand{\bibinfo}[2]{#2}
\providecommand{\BIBentrySTDinterwordspacing}{\spaceskip=0pt\relax}
\providecommand{\BIBentryALTinterwordstretchfactor}{4}
\providecommand{\BIBentryALTinterwordspacing}{\spaceskip=\fontdimen2\font plus
\BIBentryALTinterwordstretchfactor\fontdimen3\font minus
  \fontdimen4\font\relax}
\providecommand{\BIBforeignlanguage}[2]{{%
\expandafter\ifx\csname l@#1\endcsname\relax
\typeout{** WARNING: IEEEtran.bst: No hyphenation pattern has been}%
\typeout{** loaded for the language `#1'. Using the pattern for}%
\typeout{** the default language instead.}%
\else
\language=\csname l@#1\endcsname
\fi
#2}}
\providecommand{\BIBdecl}{\relax}
\BIBdecl

\bibitem{8039509}
W.~Kong, Z.~Y. Dong, Y.~Jia, D.~J. Hill, Y.~Xu, and Y.~Zhang, ``Short-term
  residential load forecasting based on {LSTM} recurrent neural network,''
  \emph{IEEE Trans. Smart Grid}, vol.~10, no.~1, pp. 841--851, Jan. 2019.

\bibitem{WANG201910}
Y.~Wang, D.~Gan, M.~Sun, N.~Zhang, Z.~Lu, and C.~Kang, ``Probabilistic
  individual load forecasting using pinball loss guided lstm,'' \emph{Appl.
  Energy}, vol. 235, pp. 10--20, Feb. 2019.

\bibitem{9917171}
Y.~Zhou, A.~S. Nair, D.~Ganger, A.~Tripathi, C.~Baone, and H.~Zhu, ``Appliance
  level short-term load forecasting via recurrent neural network,'' in
  \emph{Proc. IEEE Power Energy Soc. Gen. Meeting}, Denver, CO, USA, Jul. 2022,
  pp. 1--5.

\bibitem{WANG2021107233}
S.~Wang, X.~Deng, H.~Chen, Q.~Shi, and D.~Xu, ``A bottom-up short-term
  residential load forecasting approach based on appliance characteristic
  analysis and multi-task learning,'' \emph{Electr. Power Syst. Res.}, vol.
  196, p. 107233, Jul. 2021.

\bibitem{7894203}
S.~Singh and A.~Yassine, ``Mining energy consumption behavior patterns for
  households in smart grid,'' \emph{IEEE Trans. Emerging Top. Comput.}, vol.~7,
  no.~3, pp. 404--419, Jul.-Sep. 2019.

\bibitem{8933148}
Y.~Ji, E.~Buechler, and R.~Rajagopal, ``Data-driven load modeling and
  forecasting of residential appliances,'' \emph{IEEE Trans. Smart Grid},
  vol.~11, no.~3, pp. 2652--2661, May 2020.

\bibitem{8606124}
C.~Dinesh, S.~Makonin, and I.~V. Bajić, ``Residential power forecasting using
  load identification and graph spectral clustering,'' \emph{IEEE Trans.
  Circuits Syst. II Express Briefs}, vol.~66, no.~11, pp. 1900--1904, Nov.
  2019.

\bibitem{9102266}
------, ``Residential power forecasting based on affinity aggregation spectral
  clustering,'' \emph{IEEE Access}, vol.~8, pp. 99\,431--99\,444, May 2020.

\bibitem{chen2023graph}
X.~Chen, T.~Yu, Z.~Pan, Z.~Wang, and S.~Yang, ``Graph representation
  learning-based residential electricity behavior identification and energy
  management,'' \emph{Prot. Control Mod. Power Syst.}, vol.~8, no.~1, p.~28,
  Jun. 2023.

\bibitem{10147903}
H.~Yu, C.~Xu, G.~Geng, and Q.~Jiang, ``Multi-time-scale shapelet-based feature
  extraction for non-intrusive load monitoring,'' \emph{IEEE Trans. Smart
  Grid}, vol.~15, no.~1, pp. 1116--1128, Jan. 2024.

\bibitem{makonin2016ampds}
S.~Makonin, B.~Ellert, I.~V. Bajic, and F.~Popowich, ``{Electricity, water, and
  natural gas consumption of a residential house in Canada from 2012 to
  2014},'' \emph{Sci. Data}, vol.~3, no. 160037, pp. 1--12, Jun. 2016.

\end{thebibliography}
\end{document}